\documentclass[preprint,12pt]{aastex}
\usepackage{amsmath}
\usepackage{apjfonts}
\usepackage{graphicx}

\def\lta{\lower2pt\hbox{$\buildrel {\scriptstyle <} 
   \over {\scriptstyle\sim}$}}
\def\gta{\lower2pt\hbox{$\buildrel {\scriptstyle >} 
   \over {\scriptstyle\sim}$}}

\begin{document}

\title{Distribution of Gamma-ray Burst Ejecta Energy with Lorentz Factor}

\author{Jonathan Granot \\
   {\rm KIPAC, P.O. Box 20450, Mail Stop 29, Stanford, CA 94309\\
   Pawan Kumar \\
Department of Astronomy, University of Texas, Austin, TX 78712\\ }}

\bigskip

\begin{abstract}

The early X-ray afterglow for a significant number of gamma-ray bursts
detected by the {\it Swift} satellite is observed to have a phase of
very slow flux decline with time ($F_\nu\propto t^{-\alpha}$ with
$0.2\lesssim\alpha\lesssim 0.8$) for $10^{2.5}\;{\rm s}\lesssim
t\lesssim 10^4\;$s, while the subsequent decline is the usual
$1\lesssim\alpha_3\lesssim 1.5$ behavior, that was seen in the
pre-{\it Swift} era.
We show that this behavior is a natural consequence of a small spread
in the Lorentz factor of the ejecta, by a factor of $\sim 2-4$, where
the slower ejecta gradually catch-up with the shocked external medium,
thus increasing the energy of forward shock and delaying its
deceleration. The end of the ``shallow'' flux decay stage marks the
beginning of the Blandford-McKee self similar external shock evolution. 
This suggests that most of the energy in the relativistic outflow is in
material with a Lorentz factor of $\sim 30-50$.

\end{abstract}


\section{Introduction}

Among the discoveries made by the {\it Swift} satellite within a few
months of its launch is the observation that a fraction of long
duration gamma-ray bursts (GRBs) go through an early phase of
relatively slow decline in the X-ray afterglow flux that typically
starts at a few minutes after the burst and lasts for about an hour
\citep{Nousek05}. This phase is followed by a somewhat faster and more
typical flux decay that satisfies the expected relation between the
temporal decline index $\alpha$ and the spectral index $\beta$, where
$F_\nu\propto \nu^{-\beta} t^{-\alpha}$, similar to what was observed
before the {\it Swift} era when the monitoring of the afterglow light
curves started at least several hours after the GRB.  The spectral
index does not seem to undergo any change when the light-curve
transitions (at $t_{\rm break,2}\sim 10^4\;$s) from a shallow decline
($\alpha_2$) to the ``regular'' decline ($\alpha_3$).  It has been
argued convincingly by a number of authors that the more slowly
declining lightcurve, like the ``regular'' flux decay rate that
follows it, are both produced by the shock heated circum-burst medium
\citep{Nousek05,Panaitescu05,Zhang05}.  The
shallow X-ray flux decay is widely attributed to energy injection into
the afterglow shock, which may be caused by either a long lived
activity of the central source, or a short lived central explosion
that produces ejecta with some distribution of Lorentz factor (LF),
cf. \citep{Nousek05,Panaitescu05,Zhang05}.  In either of these
scenarios the deceleration of the afterglow shock is reduced due to
the energy being added to it, and this in turn produces a slowly
declining light curve.

Slow decline of the early optical lightcurve was also reported
before the {\it Swift} era, although quite rarely, e.g. GRB~021004
\citep{Fox03}. \citet{LC03} have made a good case that the slowly
declining early optical lightcurve of GRB~021004 could naturally occur
for a stellar wind type external medium when the optical band is below
the characteristic synchrotron frequency (and above the cooling break
frequency). However, this explanation is unlikely to work for the
slowly declining X-ray lightcurves since the X-ray band at $\sim
1\;$hr after a GRB is expected to lie well above the synchrotron
characteristic frequency, and furthermore there is no evidence for a
change in the spectral slope across the break in the X-ray lightcurve
at $t_{\rm break,2}$. \citet{Fox03} argued that the early flat optical
lightcurve of GRB~021004 is either due to energy injection into the
afterglow shock, or due to angular inhomogeneity \citep[``patchy
shell'';][]{KP00}. The latter was favored as it also provided a good
explanation for the fluctuations (or ``bumps'') that appear later in
the optical lightcurve of GRB~021004 \citep[see
also][]{NPG03}. However, the sparse early afterglow data in the
pre-{\it Swift} era made it difficult to distinguish between the
different explanations.

A long lived activity of the central source is not very appealing
since it would require the source to be active up to several hours
after the GRB, with a very smooth temporal behavior, where most of the
energy is in the outflow that is ejected around $t_{\rm break,2}\sim
10^4\;$s \citep[see however][]{Dai04}; this makes the problem of the
observed high efficiency for converting kinetic energy to gamma-ray
radiation much worse \citep{Nousek05}.  Another interesting way to
produce an early flat phase in the afterglow light curve \citep{EG05}
is by a line of sight that is slightly outside the (sharp) edge of a
roughly uniform jet \citep{Granot02,GR-RP05}. This would, however,
naturally be accompanied by a weaker and softer prompt emission,
perhaps resulting in an X-ray flash or X-ray rich GRB rather than a
classical GRB; the more pronounced this effect is the flatter and
longer lived the slow X-ray afterglow decay phase should be. Initial
inspection of the data does not show such a correlation, suggesting
that viewing angle effects are probably not the predominant cause of
the early slow decay phase in the X-ray afterglows, at least under the
simplest assumptions.\footnote{\citet{EG05} point out that viewing
angle effects might still be the dominant cause of the flat early
decay of the afterglow light curves if along some lines of sight the
kinetic energy in the afterglow shock is very low while the energy in
gamma-rays remains high.}

It is natural to expect that matter ejected in any explosion will have
a range of velocities or LFs. After a while (on a time scale, in the
observer frame, of order a few times the duration of the central
engine activity) the ejecta will rearrange themselves such that the
fastest moving plasma is at the head of the outflow and the slowest at
the tail end. This can occur either through internal shocks within the
outflow, or by a smooth decrease in the LF of the outflow toward the
end of the central source activity.  If the ejecta have a finite range
of LFs, the slower ejecta would gradually catch up with the shocked
external medium, injecting energy into the forward shock. If the
slower ejecta carry more energy than the faster ejecta, then this
added energy would gradually increase the energy of the afterglow
shock, causing it to decelerate more gradually. Once the energy in the
lower LF ejecta becomes small compared to the energy already in the
afterglow shock, the blast wave evolution becomes impulsive (i.e. the
subsequent small amount of energy injection hardly effects the
evolution of the forward shock), and if radiative losses are
unimportant then it approaches the adiabatic \citet{BM76} self-similar
solution. This occurs when the LF of the afterglow shock drops
slightly below $\Gamma_{\rm peak}$, the LF where $dE/d\ln\Gamma$ peaks
and where most of the energy in the outflow resides.

In this paper we use the {\it Swift} data to determine the time
dependence of the blast wave LF. We find that the LF typically drops
by a factor of $\sim 2-4$ during the shallow decline phase. This is
consistent with the basic picture suggested above, where a finite LF
distribution for the ejecta causes a more gradual decline of the
forward-shock LF, which gives rise to a shallow light-curve, and is an
intermediate transition stage before the onset of the adiabatic
Blandford-McKee solution.

\section{Dependence of Burst Kinetic Energy on Lorentz Factor}

The emission from an external shock can be described in terms of the
shock front LF ($\Gamma$) and the density profile of the
circum-stellar medium (CSM).  For a uniform CSM the synchrotron
characteristic frequency ($\nu_m$), the cooling frequency ($\nu_c$)
and the flux at the peak of the spectrum ($F_{\rm\nu,max}$), in the
observer frame, are proportional to $\Gamma^4$, $\Gamma^{-4} t^{-2}$
and $t^3 \Gamma^8$ respectively, where $t$ is the observed time. The
flux at a frequency between the $\nu_m$ and $\nu_c$ is proportional to
$t^3 \Gamma^{6+2p}$ and for the observed band above $\nu_m$ and
$\nu_c$ the flux scales as $t^2 \Gamma^{4+2p}$. The observed flux is
strongly dependent on $\Gamma$ and therefore even a small deviation
from the $\Gamma \propto t^{-3/8}$ scaling has a very large effect on
the observed light-curve. The observed flux has a weaker dependence on
$\Gamma$ for a wind like density stratification of the CSM; the flux
in the two regimes considered above scales roughly as
$\Gamma^{1+p}t^{(1-p)/2}$ and $\Gamma^{2+p}t^{-(p-2)/2}$,
respectively.

More generally, for a power law external density profile, $\rho_{\rm
ext}=Ar^{-k}$, we have $F_{\rm\nu,max}\propto\Gamma B
R^{3-k}\propto\Gamma^2 R^{3-3k/2}\propto \Gamma^{8-3k}t^{3-3k/2}$,
$\nu_m\propto\Gamma B\gamma_m^2\propto\Gamma^4
R^{-k/2}\propto\Gamma^{4-k}t^{-k/2}$, $\gamma_c\propto 1/\Gamma B^2 t$
and $\nu_c\propto\Gamma B\gamma_c^2
\propto\Gamma^{-1}B^{-3}t^{-2}\propto
R^{3k/2}\Gamma^{-4}t^{-2}\propto\Gamma^{3k-4}t^{-2+3k/2}$.  Therefore,
\begin{equation}
F_\nu \approx \left\{ \begin{matrix}
F_{\rm\nu,max}(\nu/\nu_c)^{-1/2} &\propto &
\Gamma^{6-3k/2}t^{2-3k/4} & 
\nu_c<\nu<\nu_m \ ,\cr \cr
F_{\rm\nu,max}(\nu/\nu_m)^{(p-1)/2} &\propto &
\Gamma^{}t^{3-k(p+5)/4} & 
\nu_m<\nu<\nu_c \ ,\cr \cr
F_{\rm\nu,max}(\nu_c/\nu_m)^{(p-1)/2}(\nu/\nu_c)^{-p/2} &\propto &
\Gamma^{4-k+p(4-k)/2}t^{2-k(2+p)/4} &
\nu>\max(\nu_m,\nu_c) \ .
\end{matrix} \right.
\end{equation}

Assuming that the LF distribution for the ejecta is
$E(>\Gamma)\propto\Gamma^{-a}$, we find $g\equiv-d\log\Gamma/d\log t$
is smaller by an amount $\delta$ compared to the standard value of
$(3-k)/2(4-k)$, i.e. 3/8 (1/4) for a uniform (wind) CSM, where
\begin{equation}
\delta = \frac{(3-k)a}{2(4-k)\left[2(4-k)+a\right]}
= \left\{\begin{matrix} 
3a/[8(8+a)] & \ \  k=0 \ , \cr\cr
a/[4(4+a)]  & \ \  k=2 \ .
\end{matrix}\right.
\end{equation}
The deviation to the LC temporal power-law index ($\Delta\alpha$) from
the standard case of Blandford-McKee self-similar solution ($\alpha$)
is easily related to $\delta$. For $\nu_m<\nu<\nu_c$ we have
\begin{equation}
\Delta\alpha = \left[6-\frac{p(4-k)-5k}{2}\right]\delta =
\left\{ \begin{matrix}
3(3+p)a/[4(8+a)] & \ \ k=0 \ , \cr\cr
(1+p)a/[4(4+a)]  & \ \ k=2 \ ,
\end{matrix}\right.
\end{equation}
while for
$\nu>\max(\nu_m,\nu_c)$, 
\begin{equation}
\Delta\alpha = \left[4-k+\frac{p(4-k)}{2}\right]\delta =
\left\{ \begin{matrix}
3(2+p)a/[4(8+a)] & \ \ k=0 \ , \cr\cr
(2+p)a/[4(4+a)]  & \ \ k=2 \ .
\end{matrix}\right.
\end{equation}

We next calculate $\delta$ for a number of Swift detected GRBs with a
shallow LC using the observed spectral index and the change in the
temporal power-law index for the X-ray lightcurve ($\Delta\alpha$)
between the shallow and the ``regular'' parts of the LC. The results
for $\delta$, and the change to the LF during the shallow LC are shown
in Table 1.

We note that the change to $\Gamma$ during the shallow phase of the LC
was calculated using the appropriate dependence of $\Gamma$ on $t$;
for a uniform CSM this is $t^{\delta-3/8}$. It can be seen in Table 1
that $\Gamma$ changes by a factor $\sim 2-4$ for all the bursts, with
a uniform CSM, during the shallow LC phase; these numbers change only
by a small amount even if we take the forward shock emission, and the
shallow decline, to begin at the end of the GRB.

The function $dE/d\ln\Gamma$ peaks at $\Gamma_{\rm
peak}\sim\Gamma(t_{\rm break,2})$, the LF of the forward shock at the
end of the shallow decline phase of the X-ray lightcurve.  For
$\Gamma>\Gamma_{\rm peak}$, $dE/d\ln\Gamma\propto \Gamma^{-a}$. The
power-law index $a$ is given in table 1 for a number of bursts detected
by Swift and lies between $\sim 1$ and $\sim 2.5$ if the CSM has
uniform density \citep[][report similar values -- $s-1$ in their
notation]{Nousek05}; $a\gta5$ if the medium in the vicinity of GRB is
taken to be a wind-CSM\footnote{The total amount of energy injection
during the shallow decline phase is independent of the stratification
of the circum-stellar medium. However, for a wind-like CSM $a\gta5$ which
means that $dE/d\ln\Gamma$ must have a very narrow peak of width
$\delta\ln\Gamma\ll1$ which is unlikely to be realized in nature.}
or alternatively the central source has to be
active for several hours with little variability and a roughly
constant rate of energy output in relativistic outflow -- neither of
these possibilities seem very plausible and so the case of a wind-CSM
is not considered any further in this paper.  For $\Gamma<\Gamma_{\rm
peak}$ the function $dE/d\ln\Gamma \propto \Gamma^b$ should decrease
with decreasing $\Gamma$ (i.e. $b>0$) as otherwise slower moving
ejecta will continue to add substantial amount of energy to the
forward shock thereby retarding its deceleration and slowing down the
decline of the lightcurve. Since the spectral index and the lightcurve
power-law decay index after the end of the shallow decline phase obey
the relationship expected for an adiabatic forward shock evolution we
conclude that indeed $b>0$, but its exact value is otherwise
unconstrained.  Radio calorimetry for a number of GRBs has concluded
that there is not a whole lot of energy in GRBs in the form of mildly
relativistic ejecta with $\Gamma\sim 2$
\citep[e.g.][]{Berger04,Frail05}. This further strengthens our
conclusion that $b>0$, and that this scaling might extends to
$\Gamma\sim 2$. We note that for a given total energy in the explosion
of order $10^{52}\;$erg the relation $dE/d\ln\Gamma \propto
\Gamma^{-a}$, with $a\sim 1.5$, must turnover at some $\Gamma$ of
order $10$ or so otherwise the energy in the relativistic ejecta will
exceed the total available energy (energy in relativistic ejecta with
$\Gamma>\Gamma_{\rm peak}\sim 50$ is of order $10^{51}\;$erg).  We
have now considerable body of evidence that long duration GRBs are
accompanied by a supernova of Type Ic, which expels a few solar masses
of material at velocities of order $10^4\;{\rm km\; s^{-1}}$.  Thus,
$dE/d\ln u$, where $u=\beta\Gamma=(\Gamma^2-1)^{1/2}$, must again
turnover over and have a peak at $u\sim 0.05$. Putting all these
together we show a schematic behavior of $E(\beta\Gamma)$
(i.e. $dE/d\ln u$) in Figure 1.

\section{Conclusion}

We have pieced together the distribution of energy in gamma-ray burst
ejecta as a function of the four-velocity
$u=\beta\Gamma=(\Gamma^2-1)^{1/2}$ for $0.1\lesssim u\lesssim
10^2$. The distribution function, $dE/d\ln u$, has two peaks: one at
$u \sim 0.1$ and another at $u\sim 30-50$. For $u\gtrsim 50$, it falls
off as $dE/d\ln u\propto u^{-a}$ with $a\sim 1-2$, as is determined
from the shallow decline of the X-ray lightcurve at early times
($10^{2.5}\;{\rm s}\lesssim t\lesssim 10^4\;$s) observed for a good
fraction of bursts detected by the Swift satellite. The distribution
at low $u\sim0.1$ is obtained by observations of supernovae Ic that
are associated with GRBs. In the intermediate regime of $1\lesssim
u\lesssim 30$ the shape of the distribution function $dE/d\ln u$ is
very uncertain, but we argue that it is likely to be at least flat or
slowly rising in this range.

A prediction of this model for the shallow decline of X-ray
light-curve, that is based on a deviation (at early times) from the
constant energy Blandford-McKee self-similar solution, is that we
should see a roughly similar shallow decline in the optical band over
the same time interval as in the X-ray data.  Since the optical and
the X-ray bands in general lie in different segments of the
synchrotron spectrum, and because the energy added to the forward
shock by slower moving ejecta should be accompanied by a mildly
relativistic reverse shock that could provide some added flux to the
optical lightcurve, the rate of decline for the X-ray and optical
lightcurves should be similar but not identical in this
model.\footnote{For the power law segments (PLSs) $\nu_m<\nu<\nu_c$
and $\nu>\max(\nu_m,\nu_c)$ the temporal decay index $\alpha$ of the
reverse shock emission is slightly smaller (corresponding to a
slightly shallower flux decay) than that for the forward shock within
the corresponding PLS, by $\Delta\alpha=g(p-2)$. For the PLSs
$\nu_c<\nu<\nu_m$ and $\nu<\nu_c<\nu_m$ the reverse shock emission
decays somewhat faster than that of the forward shock at the
corresponding PLS, by $\Delta\alpha=g$.}  A highly magnetized outflow
could significantly weaken the reverse shock (or even eliminate it
altogether) and thus suppress its emission. The alternative
explanation of a viewing angle slightly outside the edge of the jet
would lead to a gradual steepening of the afterglow lightcurve (i.e. a
gradual increase in $\alpha$, as the beaming cone of the afterglow
emission gradually approaches and eventually encompasses the line of
sight), while a steeper break in the light curve is possible (and
arguably, might also be expected) in the model described in this work,
when the stage of energy injection into the afterglow shock ends.

The challenge posed for GRB/SNe models is to understand what physical
processes give rise to $a\sim 2$ and why the LF distribution of the
ejecta peaks at a value roughly $\Gamma_{\rm peak}\sim 30-50$.
Understanding these results should help illuminate the processes
operating during the period in which the central engine of gamma-ray
burst is active and the interaction of the relativistic outflow with
the collapsing star and its immediate surroundings.

{\bf Acknowledgment: } We thank Stan Woosley for useful discussion.
This work is supported in part by grants from NASA and NSF (AST-0406878) 
to PK and by the US Department of Energy under contract number 
DE-AC03-76SF00515 (J.~G.).

\begin{deluxetable}{lcccccccccc}
\tabletypesize{\scriptsize}
\tablecaption{The Change in the Lorentz Factor During the Energy
  Injection Episode}
\tablewidth{0pt}
\tablehead{
\colhead{GRB \#} &
\colhead{$T_{90}/$s} &
\colhead{$t_{\rm break,1}/$s} &
\colhead{$t_{\rm break,2}/$s} &
\colhead{$a$} & 
\colhead{$\delta$} & 
\colhead{$\;\;\xi_{\rm min}\;^\dagger$} &
\colhead{$\xi_{\rm max}$} &
\colhead{$\Gamma_{\rm peak}\;^\star$} &
\colhead{$\Gamma_0\;^\clubsuit$}
}
\startdata
050128 & 13.8 & $<230$ & $1720^{+940}_{-570}$ & $1.1\pm 0.2$ &
$0.045\pm 0.07$ & 2.0 & 4.9 & --- & --- \\
050315 & 96.0 & $400\pm 20$ & $12000\pm 400$ & 2.4 $\pm$ 0.1 & 
$0.094\pm 0.03$ & 2.6 & 3.9 & 30 & 117 \\
050319 & 10.0 & $370\pm 15$ & $40000\pm 300$  &  $1.6\pm 0.5$ &
$0.063\pm 0.016$ & 4.3 & 13.3 & 21 & 279 \\
050401 & 33.0 & $<127$  & $5500^{+1150}_{-1050}$  &  $1.7\pm 0.1$
& $0.066\pm 0.003$ & 3.2 & 4.9 & 58 & 284 \\
050416a & 2.4 & $<80$ & $1350^{+2070}_{-620}$ & $1.1\pm 0.1$ &
$0.043\pm 0.004$ & 2.6 & 8.2 & 28 & 230 \\
050607 & 26.5 & $510\pm 50$ & $6400\pm 900$ &  $1.5\pm 0.1$ &
$0.059\pm 0.004$ & 2.2 & 5.7 & --- & --- \\
\enddata

\tablecomments{The relevant data were taken from \citet{Nousek05}. All
  of the calculated quantities reported in this table -- $a$, $\delta$,
  $\xi$ and $\Gamma_0$ -- assume a uniform density medium in the
  vicinity of these bursts. $^\dagger\;$Here $\xi$ is the ratio of the
  Lorentz factor of the afterglow shock at the start and at the end of
  the ``shallow part'' of the X-ray light-curve, and its value is
  estimated to be between $\xi_{\rm min}=(t_{\rm break,2}/t_{\rm
  break,1})^{3/(8+a)}$ and $\xi_{\rm max}=(t_{\rm
  break,2}/T_{90})^{3/(8+a)}$; $^\star\;$These values of $\Gamma_{\rm
  peak}=\Gamma(t_{\rm break,2})$ were estimated only for the GRBs with
  known redshifts, by using equation 9 of \citet{Nousek05} where the
  isotropic equivalent kinetic energy at $t_{\rm break,2}$ was taken
  to be equal to $E_{\rm\gamma,iso}$, and the external density was
  taken to be $n=1\;{\rm cm^{-3}}$; $^\clubsuit\;$The initial Lorentz
  factor is simply estimated by $\Gamma_0=\xi_{\rm max}\Gamma_{\rm
  peak}$.}
\end{deluxetable}

\vfill\eject

\begin{figure}
\includegraphics[width=15cm]{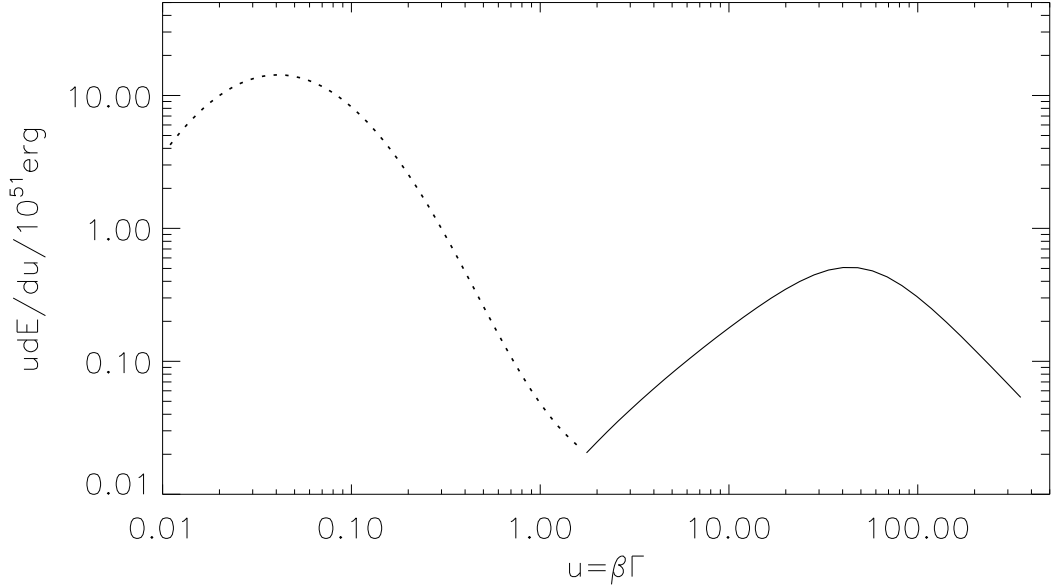}
\caption{\label{fig} Schematic figure showing $dE/d\ln u$, in units of
$10^{51}$ erg, as a function of
$u\equiv\Gamma\beta=(\Gamma^2-1)^{1/2}$.  It has one relativistic
component (solid line) with total energy $\sim 10^{51}\;$erg and peak
at $u\sim 30-50$ that produces the gamma-ray burst and the afterglow
radiations.  The power-law index above the peak for this component is
well constrained by the X-ray data (the shallow part of the
light-curve) and is $\sim -1.5$ (see table 1). The slope below the
peak is not constrained and is taken to be 1; in reality it can be
close to zero, as the only constraint we have is from late time radio
afterglow observations which suggests that there is not a lot of extra
energy in material moving with Lorentz factor of order 2. The second
component (dashed curve) shows schematically the kinetic energy in
non-relativistic ejecta in the supernova accompanying the GRB; the
peak for this component is taken to be $\sim10^4\;{\rm km\; s^{-1}}$,
the typical velocity for SNe Ic ejecta, and the energy is
$\sim10^{52}\;$erg.  }
\end{figure}


\begin{thebibliography}{}

\bibitem[Berger et al.(2004)]{Berger04}
Berger, E., Kulkarni, S. R.; Frail, D. A, 2004, ApJ 612, 966

\bibitem[Blandford \& McKee(1976)]{BM76}
Blandford, R.~D., \& McKee, C.~F. 1976, Phys. Fluids, 19, 1130

\bibitem[Dai(2004)]{Dai04}
Dai, Z.~G. 2004, ApJ, 606, 1000

\bibitem[Eichler \& Granot(2005)]{EG05}
Eichler, D., \& Granot, J. 2005, submitted to ApJL (astro-ph/0509857)

\bibitem[Fox et al.(2003)]{Fox03}
Fox, D.W. et al., 2003, Nature, 422, 284

\bibitem[Frail et al.(2005)]{Frail05}
Frail, D., Soderberg, A. M., Kulkarni, S. R., Berger, E., Yost, S., 
Fox, D. W., Harrison, F. A, 2005, ApJ 619, 994

\bibitem[Granot et al.(2002)]{Granot02}
Granot, J., Panaitescu, A., Kumar, P., \& Woosley, S.~E. 2002, ApJ,
570, L61

\bibitem[Granot, Ramirez-Ruiz \& Perna(2005)]{GR-RP05}
Granot, J., Ramirez-Ruiz, E., \& Perna, R. 2005, ApJ, 630, 1003

\bibitem[Kumar \& Piran(2000)]{KP00}
Kumar, P., \& Piran, T. 2000, ApJ, 535, 152

\bibitem[Li \& Chevalier(2003)]{LC03}
Li, Z-Y, \& Chevalier, R.A. 2003, ApJ 589, L69

\bibitem[Nakar, Piran \& Granot(2003)]{NPG03}
Nakar, E., Piran, T., \& Granot, J. 2003, New Astron., 8, 495

\bibitem[Nousek et al.(2005)]{Nousek05}
Nousek, J.~A., et al. 2005, submitted to ApJ (astro-ph/0508332)

\bibitem[Panaitescu et al.(2005)]{Panaitescu05}
Panaitescu, A., et al. 2005, submitted to MNRAS, (astro-ph/0508340)

\bibitem[Sari \& M\'esz\'aros(2000)]{SM00}
Sari, R., \& M\'esz\'aros, P. 2000, ApJ, 535, L33

\bibitem[Zhang et al.(2005)]{Zhang05}
Zhang, B., et al. 2005, submitted to ApJ (astro-ph/0508321)


\end{thebibliography}
\end{document}